\documentclass[fleqn,10pt]{wlscirep}
\usepackage[utf8]{inputenc}
\usepackage[T1]{fontenc}
\usepackage{cite}
\usepackage{amsmath,amssymb,amsfonts,mathtools}
\usepackage{multirow}
\usepackage{booktabs}
\usepackage{graphicx}
\usepackage[caption=false]{subfig}
\usepackage{cleveref}
\title{Clustering Coefficient Reflecting\\Pairwise Relationships within Hyperedges}

\author[1]{Rikuya Miyashita}
\author[2,*]{Shiori Hironaka}
\author[2]{Kazuyuki Shudo}
\affil[1]{Tokyo Institute of Technology, Department of Mathematical and Computing Science, Tokyo, 152-8552, Japan}
\affil[2]{Kyoto University, Academic Center for Computing and Media Studies, Kyoto, 606-8501, Japan}

\affil[*]{hironaka@media.kyoto-u.ac.jp}


\begin{abstract}
    Hypergraphs are generalizations of simple graphs that allow for the representation of complex group interactions beyond pairwise relationships.
    Clustering coefficients quantify local link density in networks and have been widely studied for both simple graphs and hypergraphs.
    However, existing clustering coefficients for hypergraphs treat each hyperedge as a distinct unit rather than a collection of potentially related node pairs, failing to capture intra-hyperedge pairwise relationships and incorrectly assigning zero values to nodes with meaningful clustering patterns.
    We propose a novel clustering coefficient that addresses this fundamental limitation by transforming hypergraphs into weighted graphs, where edge weights reflect relationship strength between nodes based on hyperedge connections.
    Our definition satisfies three key conditions: values in the range [0,1], consistency with simple graph clustering coefficients, and effective capture of intra-hyperedge pairwise relationships—a capability absent from existing approaches.
    Theoretical evaluation on higher-order motifs demonstrates that our definition correctly assigns values to motifs where existing definitions fail (motifs III, IV-a, IV-b of order 3), while empirical evaluation on three real-world datasets shows similar overall clustering tendencies with more detailed measurements, especially for hypergraphs with larger hyperedges.
    The proposed clustering coefficient enables accurate quantification of local density in complex networks, revealing structural characteristics missed by existing definitions in systems where group membership implies connections between members, such as social communities and co-authorship networks.
\end{abstract}

\newenvironment{DIFnomarkup}{}{}

\begin{document}

\flushbottom
\maketitle
\thispagestyle{empty}

\section*{Introduction}

A network is a structure that represents components and their interactions.
The most common type of network is the undirected simple graph, which captures pairwise interactions between components.
However, real-world networks often involve interactions among more than two nodes.
For instance, in an email network, a single email may involve multiple senders and receivers~\cite{Klimt2004}, creating a group interaction.

Hypergraphs, generalizations of undirected simple graphs, represent networks with group interactions.
In a hypergraph, an edge (called a hyperedge) can connect any number of nodes, enabling representation of complex relationships beyond pairwise interactions.
Hypergraphs have been used to model various real-world systems, such as collaboration networks~\cite{Aksoy2020}, biological networks~\cite{Zhou2011,Gallagher2013}, social networks~\cite{Yang2019}, and social tagging networks~\cite{Cattuto2007,Zhang2010}, where group interactions are prevalent.

The clustering coefficient, which quantifies relationships among three nodes, is a main statistic for graph analysis.
The clustering coefficient for undirected simple graphs~\cite{Watts1998} measures the likelihood that two neighbors of a node are also connected, forming a triangle.
It provides insights into the local link density and the tendency of nodes to cluster together in a network.
Clustering coefficients are widely used in various applications such as brain network analysis~\cite{Masuda2018},
modeling protein interactions~\cite{Gallagher2013}, generative models~\cite{Inoue2022,Behague2023}, and link prediction~\cite{Wu2016,Chen2019}.
Many clustering coefficients for hypergraphs have also been proposed~\cite{Estrada2006,Zhou2011,Gallagher2013,Aksoy2020,Kim2023,Ha2024}.

Although hypergraphs are a generalization of simple graphs, most definitions of the clustering coefficient for hypergraphs are inconsistent with the definitions of clustering coefficients for simple graphs.
Only two definitions which are consistent with the definition for undirected simple graphs have been proposed: one by Opsahl et al.~\cite{Opsahl2013} and another by Zhou et al.~\cite{Zhou2011}.
The challenge lies in quantifying neighborhood connectivity in hypergraphs while maintaining consistency with definitions for simple graphs.
Opsahl et al.\ addressed this by transforming hypergraphs into bipartite graphs, while Zhou et al.\ adopted the concept of extra overlap of hyperedges.
Both approaches aimed to define the proportion of loops created by hyperedges.

However, these existing definitions~\cite{Opsahl2013,Zhou2011} focus exclusively on relationships between different hyperedges while completely ignoring the internal structure within each hyperedge, treating each hyperedge as a distinct unit rather than a collection of potentially related node pairs.
This fundamental limitation becomes critical in real-world applications where hyperedge membership inherently implies pairwise relationships, such as co-authorship networks where all authors naturally collaborate.
Existing definitions incorrectly assign zero clustering coefficients to nodes whose triangular relationships exist within single hyperedges (\Cref{table:Motifs}), failing to capture the semantic meaning of group membership in hypergraphs.

To address this fundamental limitation, we propose a novel definition of the clustering coefficient for hypergraphs that accurately captures local link density by using pairwise relationships within hyperedges, while also maintaining consistency with the definition for undirected simple graphs on hypergraphs consisting solely of hyperedges of size 2.
Our approach transforms hypergraphs into weighted graphs, where the edge weights reflect the strength of relationships between nodes based on their hyperedge connections.
This transformation allows for a more detailed measurement of local link density that accurately reflects intra-hyperedge relationships, which existing definitions fail to capture.
This capability is critical for properly analyzing networks where group membership implies some degree of connection between all members, such as social communities or co-authorship relationships.
By explicitly modeling the strength of relationships between all pairs of nodes that share a hyperedge, our definition provides a more comprehensive and intuitive measure of clustering in hypergraphs, directly addressing the shortcomings of existing definitions.

The key contributions of this work are as follows:
\begin{enumerate}
\item We introduce a novel clustering coefficient that effectively captures pairwise relationships within hyperedges by transforming hypergraphs into weighted undirected graphs. This transformation preserves essential structural information while enabling more detailed measurements of local clustering tendencies.
\item We provide theoretical validation of our approach through analysis of higher-order motifs, demonstrating that our definition satisfies important properties while overcoming limitations of existing metrics.
\item Our empirical evaluation on diverse real-world hypergraph datasets confirms the practical utility of our approach, particularly for hypergraphs with large hyperedges where traditional definitions often produce extreme values.
\end{enumerate}

The proposed clustering coefficient has potential applications in various domains where hypergraph representations naturally arise, such as co-authorship networks, protein interaction networks, and social media communities. By more accurately measuring the local density of connections, our approach can reveal structural characteristics and patterns that existing definitions might miss, particularly in networks dominated by large group interactions.

\section*{Related Work}

The clustering coefficient quantifies the degree to which nodes in a network tend to cluster together.
There are two types of clustering coefficients: local clustering coefficient and global clustering coefficient~\cite{Watts1998,Newman2018}.
The local clustering coefficient is calculated for each node, while the global clustering coefficient is calculated for the entire network.
This paper focuses on local clustering coefficients for hypergraphs that are consistent with the definition for simple graphs.

Various clustering coefficients for hypergraphs have been proposed.
Zhou et al.~\cite{Zhou2011} defined the local and global clustering coefficients for hypergraphs in a manner consistent with the definition for undirected simple graphs.
Gallagher et al.~\cite{Gallagher2013} introduced several clustering coefficients for nodes and node pairs in hypergraphs and investigated their physical interpretation in the context of protein interactions.
Aksoy et al.~\cite{Aksoy2020} characterized walks on hypergraphs using $s$-walks and defined local clustering coefficients and global clustering coefficients of order $s$.
Kim et al.~\cite{Kim2023} defined transitivity, which corresponds to the clustering coefficient, for hypergraphs at both the hyperwedge and the hypergraph levels.
Ha et al.~\cite{Ha2024} proposed a clustering coefficient for hypergraphs based on quads, which are the shortest closed paths when a hypergraph is transformed into a bipartite graph.
Among these definitions, only Zhou et al.'s definition yields values matching the clustering coefficient values on simple graphs when calculated on hypergraphs consisting solely of size-2 hyperedges.

Hypergraphs can be transformed into bipartite graphs, and clustering coefficients can be calculated on the resulting bipartite graphs.
Several clustering coefficients for bipartite graphs have been proposed, including those defined based on 4-paths on bipartite graphs~\cite{Robins2004,Lind2005,Zhang2008,Aksoy2017}, degree of overlap of neighboring nodes for a pair of nodes~\cite{Latapy2008}, and paths of length 6~\cite{Opsahl2013}.
When the original hypergraph can be represented as a simple graph, only Opsahl's definition~\cite{Opsahl2013} yields values that match the clustering coefficient values of the simple graph.

To the best of our knowledge, among the existing clustering coefficients for hypergraphs and bipartite graphs, only the definitions proposed by Zhou et al.~\cite{Zhou2011} and Opsahl~\cite{Opsahl2013} yield definitions that are consistent with the clustering coefficient of undirected simple graphs.
\Cref{table:Motifs} shows the results of calculating these definitions on all the hypergraphs of order 3.
However, as evident from this table, these definitions assign a value of 0 to motifs III, IV-a, and IV-b, despite the existence of relationships via hyperedges among the three nodes.
In this paper, we propose a clustering coefficient that utilizes the pairwise relationships within a single hyperedge, which the existing definitions fail to leverage, and that reflects local density.

The clustering coefficient, as a fundamental structural metric, plays a significant role in characterizing local connectivity patterns that influence learning outcomes.
Zhou et al.~\cite{Zhou2006} pioneered the application of hypergraph structural properties in machine learning contexts, particularly for spectral clustering and semi-supervised learning tasks.
Building upon this foundation, Agarwal et al.~\cite{Agarwal2006} demonstrated how higher-order metrics in hypergraphs can enhance feature selection and dimensionality reduction in classification problems.
More recently, Sheikhpour et al.~\cite{Sheikhpour2025} developed a novel approach using hypergraph Laplacian-based semi-supervised discriminant analysis for sparse feature selection, demonstrating how structural properties such as clustering coefficients can be leveraged to improve learning performance in high-dimensional spaces.
The relationship between clustering coefficients and learning performance has been further explored by Liu et al.~\cite{Liu2021}, who showed that localized structural properties of hypergraphs can improve both clustering accuracy and computational efficiency in large-scale applications.
Thus, effective hypergraph modeling has found applications across diverse fields, and the proposal of novel clustering coefficients holds promise for advancing these domains.

\section*{Preliminaries}

We represent a simple hypergraph as $G=(V, E)$, where $V = \{v_1, v_2, \dots, v_N\}$ is the node set and $E =\{e_1, e_2, \dots, e_M\}$ is the hyperedge set.
Each element $e_i \in E$ is a subset of the node set $V$ (i.e., $e_i \subseteq V$).
Here, $N$ is the number of nodes and $M$ is the number of hyperedges.
The size of a hyperedge is defined as the number of nodes that belong to it.
We assume no multiple hyperedges exist (i.e., no two hyperedges contain exactly the same node set).
As $G$ is a general hypergraph, it is considered a non-uniform hypergraph, allowing hyperedges of different sizes.

The hypergraph $G=(V, E)$ can be transformed to and from the bipartite graph $G'=(V, E, \mathcal{E})$, where the sets $V$ and $E$ in $G$ are the two node sets of the bipartite graph, and the set $\mathcal{E}$ is the edge set of the bipartite graph.
An edge $(v_i, e_j)$ exists if and only if $v_i$ belongs to the hyperedge $e_j$ in the hypergraph.

\section*{Proposed Clustering Coefficient} \label{sec:propose}

\begin{table*}[t]
    \caption{Clustering coefficients for higher-order motifs of order 3. The existing definitions fail to properly measure the values of motifs III and IV, whereas the proposed definition can assign values to them.}
    \label{table:Motifs}
    \centering
    \begin{tabular}{|l|c|c|c|c|c|c|c|}
    \hline
    & I & II & III & IV-a & IV-b & V & VI \\ \hline
    &
    \begin{minipage}{1.2cm}
    \centering\includegraphics[width=\linewidth]{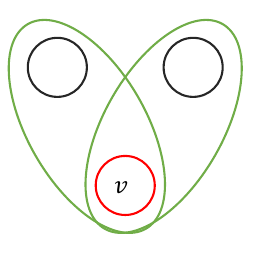}
    \end{minipage}
    &
    \begin{minipage}{1.2cm}
    \centering\includegraphics[width=\linewidth]{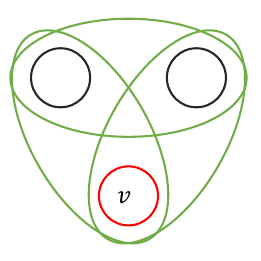}
    \end{minipage}
    &
    \begin{minipage}{1.2cm}
    \centering\includegraphics[width=\linewidth]{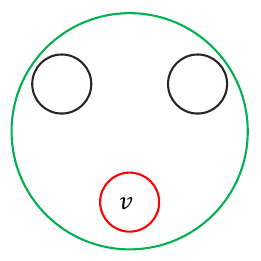}
    \end{minipage}
    &
    \begin{minipage}{1.2cm}
    \centering\includegraphics[width=\linewidth]{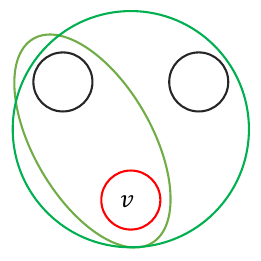}
    \end{minipage}
    &
    \begin{minipage}{1.2cm}
    \centering\includegraphics[width=\linewidth]{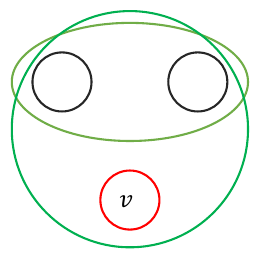}
    \end{minipage}
    &
    \begin{minipage}{1.2cm}
    \centering\includegraphics[width=\linewidth]{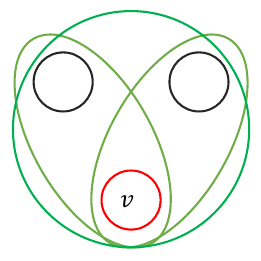}
    \end{minipage}
    &
    \begin{minipage}{1.2cm}
    \centering\includegraphics[width=\linewidth]{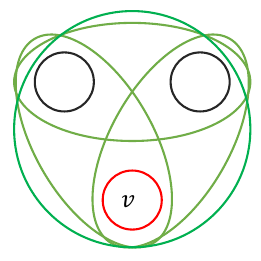}
    \end{minipage}
    \\ \hline
    $C_{\mathrm{Opsahl}}(v)$~\cite{Opsahl2013} & 0 & 1 & 0 & 0 &  0 & 1/3 & 1 \\ \hline
    $C_{\mathrm{Zhou}}(v)$~\cite{Zhou2011} & 0 & 1 & 0 & 0 & 0 & 1/3 & 1/3 \\ \hline
    $C_{\mathrm{baseline}}(v)$ & 0 & 1 & 1 & 1 & 1 & 1 & 1 \\ \hline
    $C_{\mathrm{proposed}}(v)$ & 0 & 1 & \textbf{1/2} & \textbf{1/2} & \textbf{1} & 1/2 & 1 \\ \hline
    \end{tabular}
\end{table*}

We propose a novel local clustering coefficient for hypergraphs that captures local link density using pairwise relationships within hyperedges.
The proposed clustering coefficient measures the degree of connection among neighboring nodes in a hypergraph.
The key idea is transforming the hypergraph into a weighted undirected graph, then calculating the local clustering coefficient on the resulting graph.
This approach allows for a more detailed reflection of the pairwise relationships between nodes.
As the proposed coefficient approaches 0, it indicates weak connections between a node and its neighbors.
Conversely, as it approaches 1, it signifies strong connections among the neighbors.

The proposed clustering coefficient satisfies the following three conditions:\nobreak
\begin{enumerate}
    \item The values of the proposed clustering coefficient fall within the range $[0, 1]$.
    \item The proposed clustering coefficient is consistent with the clustering coefficient of undirected simple graphs~\cite{Watts1998}.
    \item The proposed clustering coefficient effectively captures pairwise relationships within a hyperedge.
\end{enumerate}
Conditions 1 and 2 align with those outlined by Zhou et al.~\cite{Zhou2011}, while Condition 3 distinguishes our definition from existing approaches.

\begin{figure}[tp]
    \centering
    \includegraphics[width=0.5\linewidth]{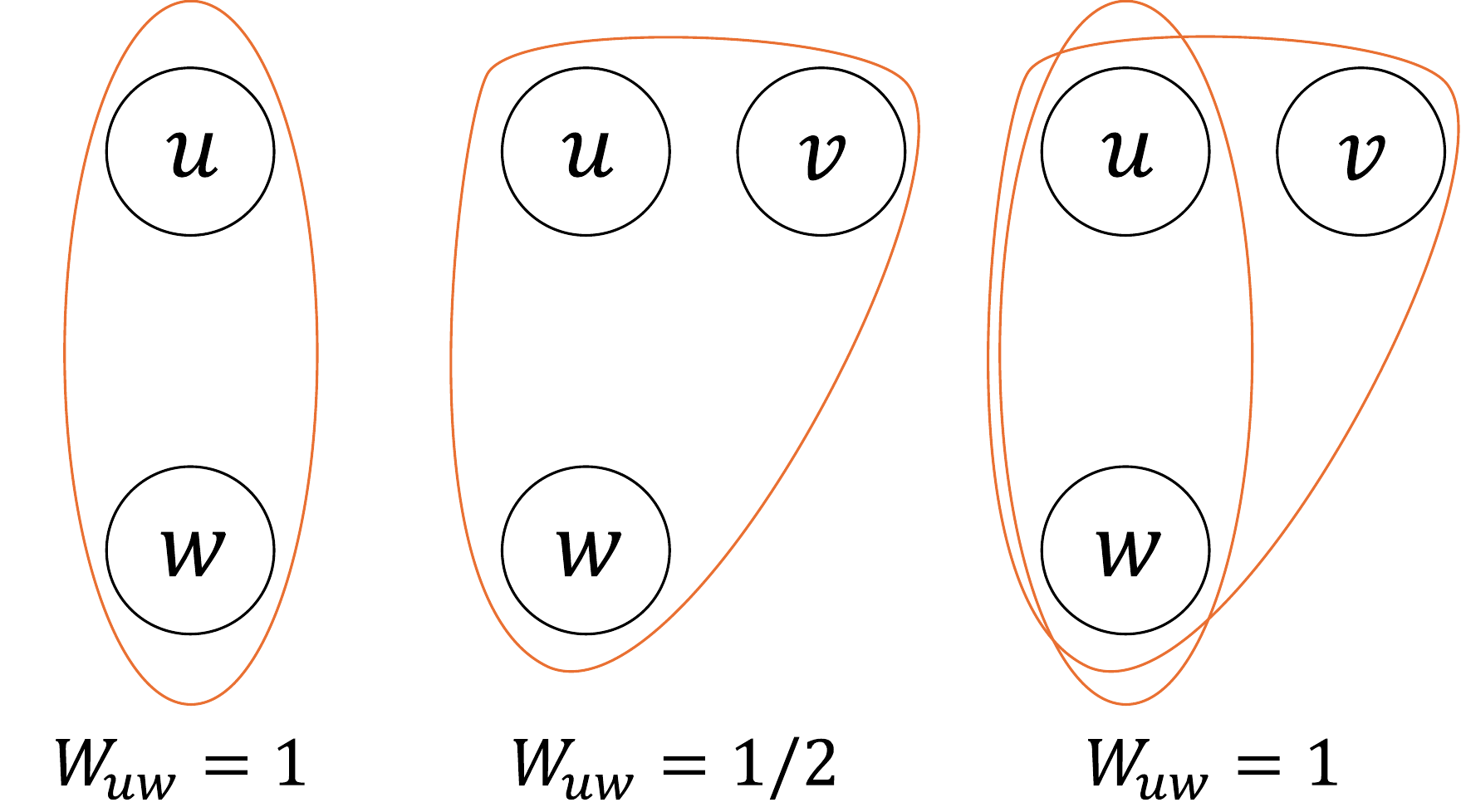}
    \caption{Examples of the proposed edge weight between nodes $u$ and $w$. The weight of edge $W_{uw}$ is determined by the maximum size of the connecting hyperedge between nodes $u$ and $w$.}
    \label{figure:weight}
\end{figure}

The transformation converts the hypergraph into a weighted undirected graph, where the edge weight $W_{vw}$ between nodes $v$ and $w$ is determined by the maximum size of the connecting hyperedge between $v$ and $w$.
The weight is maximized when the hyperedge contains only $v$ and $w$, and decreases as the hyperedge size increases.
When there is no hyperedge containing $v$ and $w$ together, the weight $W_{vw}$ is $0$.
Formally, the edge weight is defined as:
\begin{equation}
    W_{vw} = \begin{dcases}
        \max_{e \in E} \frac{1}{|e| - 1} & \text{if $\{ v, w \} \subseteq e, v \neq w$,} \\
        0 & \text{otherwise.}
    \end{dcases}
\end{equation}
\Cref{figure:weight} illustrates this definition of edge weight using examples.

The edge weight $W_{vw}$ can be interpreted as follows: the term $1/(|e|-1)$ represents the probability of randomly selecting node $w$ from the remaining nodes in hyperedge $e$ when starting from node $v$.
This captures the likelihood of a random pairwise interaction within the hyperedge. The max operation selects the strongest such connection across all hyperedges containing both nodes, ensuring that the most significant relationship is preserved in the transformation.

While this weight contains heuristic elements, it was designed to satisfy three key requirements: (1) weights must fall within $[0,1]$, (2) connections in smaller hyperedges receive higher weights than those in larger hyperedges, reflecting stronger relationships in more intimate groups, and (3) for hypergraphs consisting only of size-2 hyperedges, all weights equal 1, ensuring consistency with simple graphs.
Alternative weights which satisfy these requirements could certainly be explored. However, exploring alternative weights will be considered as future work, as our primary contribution is the framework for capturing pairwise relationships within hyperedges rather than the specific weight itself.

We then define the weight of a triangle formed by a target node $v$ and its neighbors $u$ and $w$ as the product of the edge weights: $W_{uv} \cdot W_{vw} \cdot W_{uw}$.
This formulation ensures that the triangle weight is zero if any of the edge weights is zero, effectively capturing the presence and strength of the triangular relationship.

To establish a basis for the calculation, we define the sum of the weights of the triangles that may exist among the target node $v$ and its neighbors.
This basis corresponds to the maximum number of edges that can exist between the neighbors in the context of the clustering coefficient for undirected simple graphs.
To achieve this, we define the basis as $W_{uv} \cdot W_{vw} \cdot 1$, where the weight between the neighbors $u$ and $w$ is set to $1$. This definition allows us to compare the actual weight of the triangles formed by the target node and its neighbors to the potential weight of the triangles, assuming the neighbors are fully connected with a weight of $1$.

The proposed clustering coefficient $C_{\mathrm{proposed}}(v)$ for node $v$ is then calculated as the ratio of the sum of actual triangle weights to the sum of potential triangle weights:
\begin{equation}
    C_\mathrm{proposed}(v) = \frac{ \sum\limits_{i, j \in N(v), i \neq j} W_{iv} \cdot W_{vj} \cdot W_{ij} }{ \sum\limits_{i, j \in N(v), i \neq j} W_{iv} \cdot W_{vj} \cdot 1 }
\end{equation}
where $N(v)$ denotes the set of neighbors of $v$ in the transformed weighted graph.
This calculation follows the definition of the clustering coefficient on the weighted graph~\cite{Zhang2005}. %
This approach presents a straightforward definition suitable for weighted graphs where all weights fall within the range $[0, 1]$, which is precisely the case in our graph derived from the hypergraph conversion.
As all edge weights lie within the range $[0, 1]$, the resulting clustering coefficient is guaranteed to fall within the same range, thus satisfying condition 1.

The computational complexity of our proposed definition consists of two parts: constructing the weighted graph and calculating the clustering coefficient.
The weighted graph construction requires $O(\sum_{e \in E} |e|^2)$ time, which is $O(M s^2)$ where $M$ is the number of hyperedges and $s$ is the average hyperedge size.
The clustering coefficient calculation for all nodes requires $O(\sum_{v \in V} d(v)^2)$ time, which is $O(Nd^2)$ where $N$ is the number of nodes and $d$ is the average degree in the weighted graph.

\section*{Evaluation}

We conduct theoretical and empirical evaluations of our proposed clustering coefficient.
First, we evaluate the proposed clustering coefficient using higher-order motifs of order 3.
Next, we evaluate the proposed clustering coefficient using real-world hypergraph datasets.
We thus demonstrate that our proposed clustering coefficient can calculate values reflecting pairwise relationships within hyperedges.

\subsection*{Comparative Definitions}

We compare our proposed definition of hypergraph clustering coefficients with two existing definitions that are consistent with the clustering coefficients for undirected simple graphs~\cite{Watts1998}: Opsahl's definition~\cite{Opsahl2013} and Zhou et al.'s definition~\cite{Zhou2011}.
Both definitions yield values between 0 and 1, and they are consistent with the clustering coefficients for simple graphs when applied to hypergraphs containing only size-2 hyperedges.

Opsahl's clustering coefficient~\cite{Opsahl2013} is designed for bipartite graphs and can be calculated on a hypergraph treated as a bipartite graph.
It is defined as follows:
\begin{equation}
C_{\mathrm{Opsahl}}(v) = \frac{\tau_{v,\Delta}^*}{\tau_v^*}
\end{equation}
where $\tau_v^*$ is the number of 4-paths centered on node $v$, and $\tau_{v,\Delta}^*$ is the subset of these in which the first and the last nodes of the path share a common node that is not part of the 4-path.
To count closed 4-paths on a bipartite graph, this definition counts the number of 6-paths on the bipartite graph that form cycles centered on $v$, thereby forming triangles in the hypergraph representation.

Zhou et al.~\cite{Zhou2011} propose a clustering coefficient that reflects the extent of connectivity among neighbors of node $v$ due to hyperedges other than those connecting $v$ with those neighbors.
They utilize the extra overlap between hyperedges containing node $v$, defined as follows:
\begin{align}
    C_{\mathrm{Zhou}}(v) &= \begin{dcases}
        \frac{1}{\binom{|M(v)|}{2}}\sum_{e_i, e_j\in M(v)} \mathit{EO}(e_i, e_j) & \text{if $|M(v)| > 1$} \\
        0 & \text{if $|M(v)| = 1$}
    \end{dcases}\\
    \mathit{EO}(e_i, e_j) &= \frac{|N(D_{ij})\cap D_{ji}|+|N(D_{ji})\cap D_{ij}|}{|D_{ij}|+|D_{ji}|}
\end{align}
where $M(v)$ is the set of hyperedges containing node $v$, $N(U)$ is the intersection of sets $N(v) = \{ u \mid u \in V, e \in E, \{ u, v \} \subseteq e \} \cap \{ v \}$ for each $v \in U$, and $D_{ij} = e_i - e_j$.

For comparison, we convert a hypergraph to a simple graph using clique expansion, in which each hyperedge is represented as a clique in an undirected simple graph, and calculate the clustering coefficient for the simple graph~\cite{Watts1998}, as follows:
\begin{equation}
    C_\mathrm{baseline}(v) = \frac{2 \sum_u \sum_w A_{uv} A_{vw} A_{uw}}{\sum_u A_{uv} (\sum_u A_{uv} - 1)}
\end{equation}
where $A = (A_{ij})$ is the adjacency matrix with $A_{ij} = 1$ if $\{ v_i, v_j \} \subseteq e, e \in E$ and $A_{ij} = 0$ otherwise.

\subsection*{Theoretical Evaluation on the Higher-order Motifs of Order 3}

To verify the proposed definition meets Conditions 2 and 3 presented earlier in the Proposed Clustering Coefficient section, we calculate clustering coefficients on 3-node hypergraphs based on order-3 motifs~\cite{Lotito2022}.
The results are shown in \Cref{table:Motifs}.

Motifs I and II in \Cref{table:Motifs} are representable as simple graphs.
All definitions including the proposed definition match the simple graph clustering coefficients, satisfying Condition 3.

In the existing definitions, the clustering coefficient is $0$ in motifs III, IV-a, and IV-b, where nodes are involved by the hyperedges.
For the densest motif VI, $C_{\mathrm{Zhou}}$ is only $1/3$.
In contrast, $C_{\mathrm{proposed}}$ is greater than 0 when neighbors are related and $1$ for motif VI, capturing pairwise neighbor relations missed by the existing definitions and satisfying Condition 3.

\subsection*{Empirical Evaluation Using Real-world Datasets}

In this section, we demonstrates
the advantages of our proposed definition in two ways: (1) The proposed definition can measure clustering characteristics similar to existing clustering coefficients contained in hypergraphs. (2) The proposed definition takes extreme values like 0 or 1 less frequently and can calculate values according to local density where existing definitions would assign a value of 0.

\begin{figure*}[tp]
    \centering
    \subfloat[primary-school.]{\includegraphics[width=0.33\linewidth]{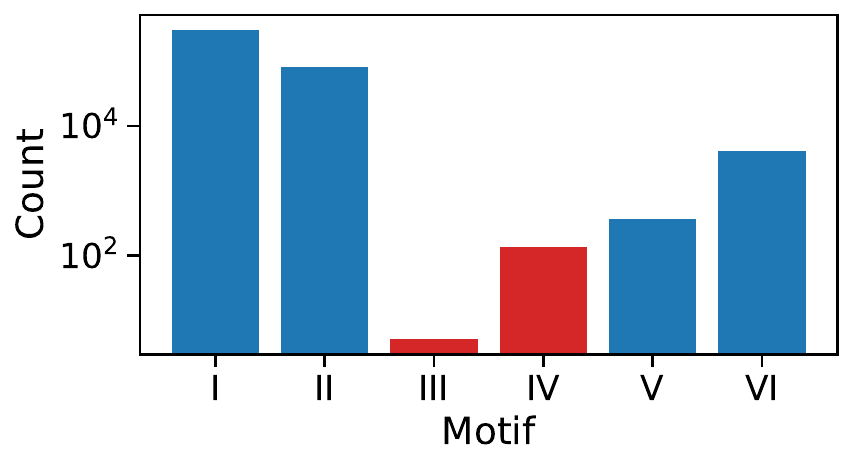}\label{fig:3a}}
    \subfloat[email-Enron.]{\includegraphics[width=0.33\linewidth]{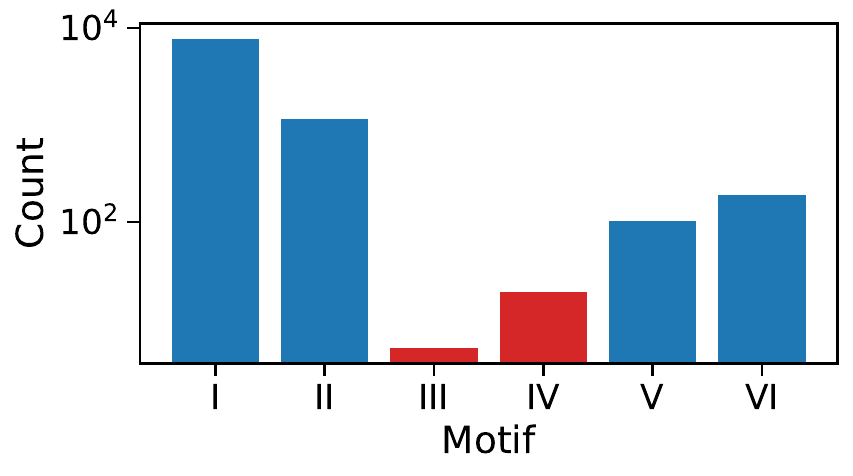}\label{fig:3b}}
    \subfloat[NDC-classes.]{\includegraphics[width=0.33\linewidth]{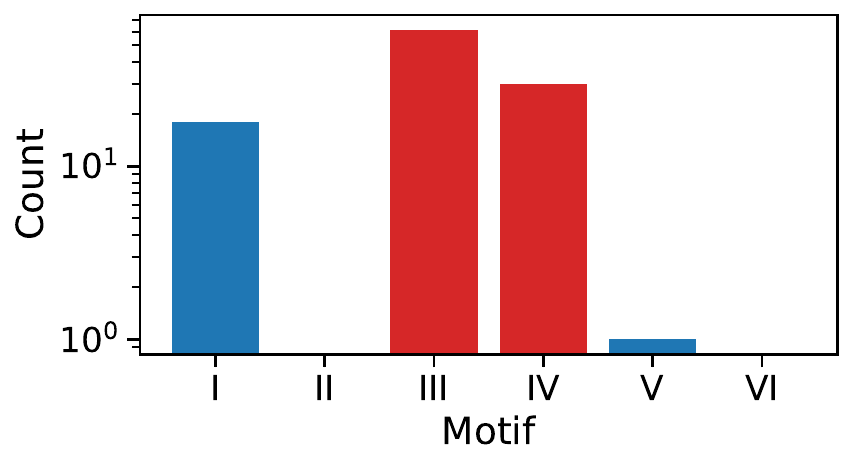}\label{fig:3c}}
    \caption{Number of higher-order motifs of order $3$ for each dataset.}
    \label{figure:Motifs}
\end{figure*}

We evaluate the proposed clustering coefficient on three real-world hypergraphs: primary-school~\cite{Benson2018,Stehle2011}, email-Enron\cite{Benson2018,Klimt2004}, and NDC-classes~\cite{Benson2018}.
The primary-school hypergraph is a contact network, which nodes represent persons and hyperedges represent sets of persons who contact each other face-to-face at close range.
The email-Enron hypergraph is an email network, which nodes represent email addresses and hyperedges represent sets of all addressees of senders and receivers of each email.
The NDC-classes hypergraph is a drug network, which nodes represent class labels (e.g., serotonin reuptake inhibitor) and hyperedges represent sets of class labels applied to each drug.
We removed multiple hyperedges from the original hypergraph and extracted the largest connected component.
We removed multiple hyperedges because our definition does not account for edge multiplicity.
We used the largest connected component to avoid disconnected subgraphs that might skew clustering coefficient calculations, since clustering coefficients are most meaningful in connected networks.
\Cref{table:Dataset} provides dataset statistics, and \Cref{figure:Motifs} shows the order-3 motif counts.
Primary-school has smaller hyperedges while NDC-classes has larger ones.
Primary-school and email-Enron have similar motif distributions, contrasting with NDC-classes which has more motifs III and IV.
In addition to these three datasets, we provide evaluation results on five additional diverse datasets in the Supplementary Information, including both small-scale (DavisClub) and large-scale networks (threads-math-sx with over 170,000 nodes)~\cite{Lotito2022}, which further demonstrate the robustness and generalizability of our approach across various network scales and domains.

\begin{table*}[tp]
    \caption{Dataset statistics. $N$: number of nodes, $M$: number of hyperedges, $\mathcal{M}$: number of edges in the corresponding bipartite graph, $\bar{k}$: average degree of the node, $\bar{s}$: average size of the hyperedge, and $\bar{C}_{\mathrm{Opsahl}}$, $\bar{C}_{\mathrm{Zhou}}$, $\bar{C}_\mathrm{proposed}$, and $\bar{C}_{\mathrm{baseline}}$ are the averages of $C_{\mathrm{Opsahl}}(v)$, $C_{\mathrm{Zhou}}(v)$, $C_\mathrm{proposed}(v)$, and $C_{\mathrm{baseline}}(v)$ for all nodes $v$.}
    \label{table:Dataset}
    \centering
    \begin{tabular}{l rrrrr ccc c}
    \toprule
    Dataset & $N$ & $M$ & $\mathcal{M}$ & $\bar{k}$ & $\bar{s}$ & $\bar{C}_{\mathrm{Opsahl}}$ & $\bar{C}_{\mathrm{Zhou}}$ & $\bar{C}_{\mathrm{proposed}}$ & $\bar{C}_\mathrm{baseline}$ \\
    \midrule
    primary-school~\cite{Benson2018,Stehle2011} & 242 & 12704 & 30729 & 126.98 & 2.42 & $0.70$ & $0.67$ & $0.51$ & 0.53 \\ 
    email-Enron~\cite{Benson2018,Klimt2004} & 143 & 1512 & 4550 & 31.82 & 3.01 & $0.68$ & $0.52$ & $0.41$ & 0.59 \\ 
    NDC-classes~\cite{Benson2018} & 628 & 816 & 5688 & 9.06 & 6.97 & $0.31$ & $0.14$ & $0.23$ & 0.77 \\ 
    \bottomrule
    \end{tabular}
\end{table*}

\begin{DIFnomarkup}
\begin{figure*}[tp]
    \centering
    \subfloat[primary-school.]{\includegraphics[width=0.33\linewidth]{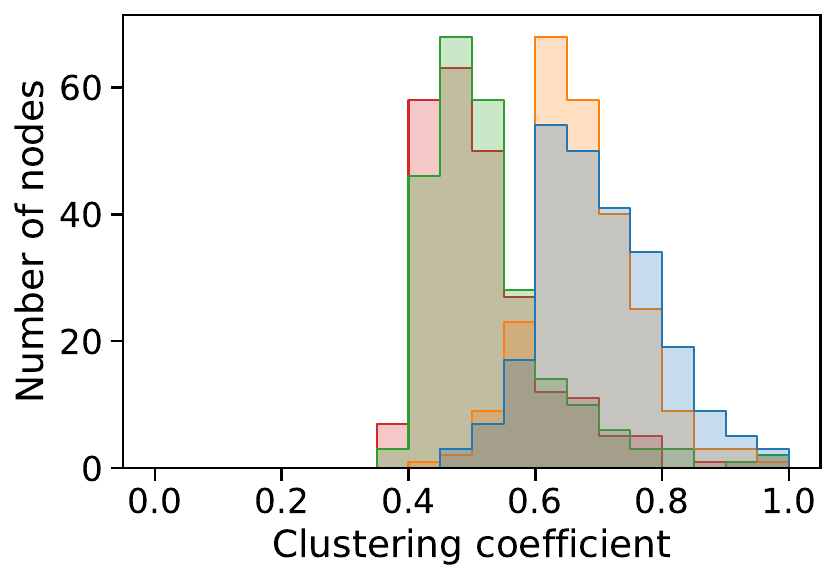}\label{fig:1a}}
    \subfloat[email-Enron]{\includegraphics[width=0.33\linewidth]{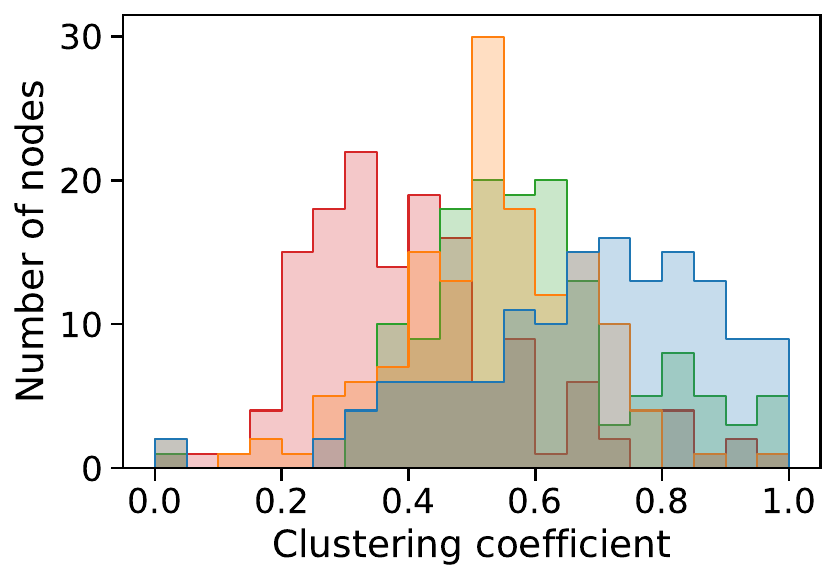}\label{fig:1b}\centering}
    \subfloat[NDC-classes.]{\includegraphics[width=0.33\linewidth]{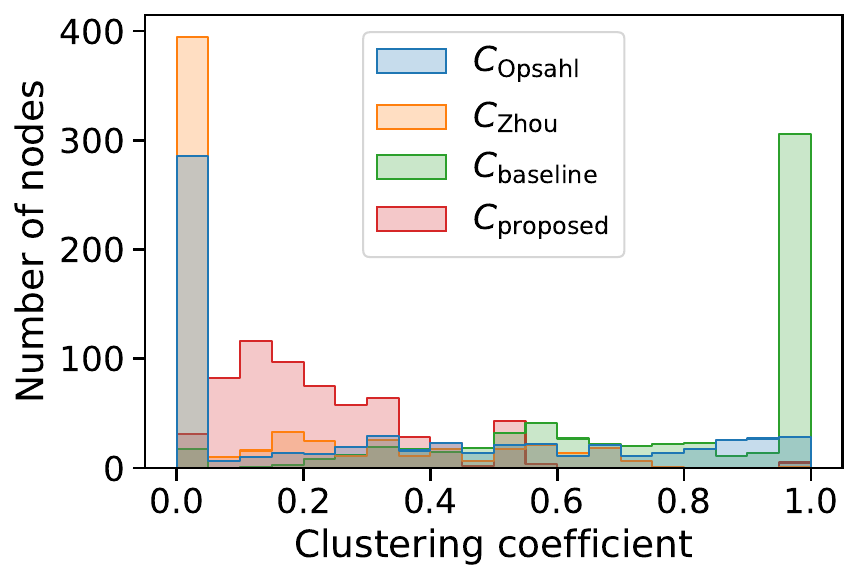}\label{fig:1c}}
    \centering
    \\
    \subfloat[primary-school]{\begin{minipage}[t]{0.333\linewidth}\centering\includegraphics[width=\linewidth]{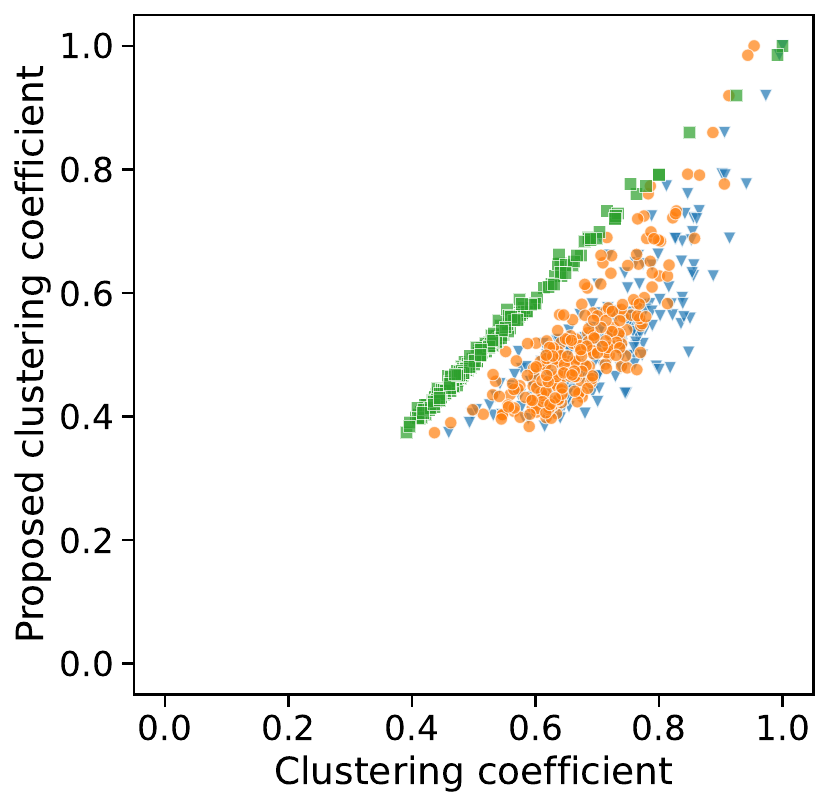}\\$\rho_{\mathrm{OP}}=0.827$, $\rho_{\mathrm{ZP}}=0.840$, $\rho_\mathrm{SP}=0.998$\end{minipage}\label{fig:2a}}
    \subfloat[email-Enron]{\begin{minipage}[t]{0.333\linewidth}\centering\includegraphics[width=\linewidth]{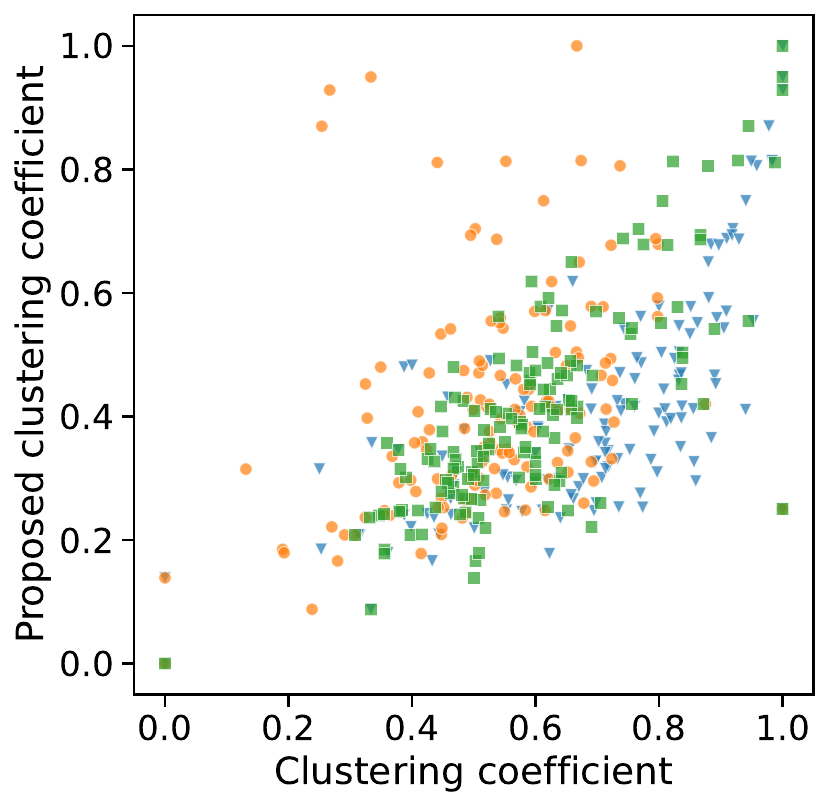}\\$\rho_{\mathrm{OP}}=0.702$, $\rho_{\mathrm{ZP}}=0.346$, $\rho_\mathrm{SP}=0.783$.\end{minipage}\label{fig:2b}}
    \subfloat[NDC-classes]{\begin{minipage}[t]{0.333\linewidth}\centering\includegraphics[width=\linewidth]{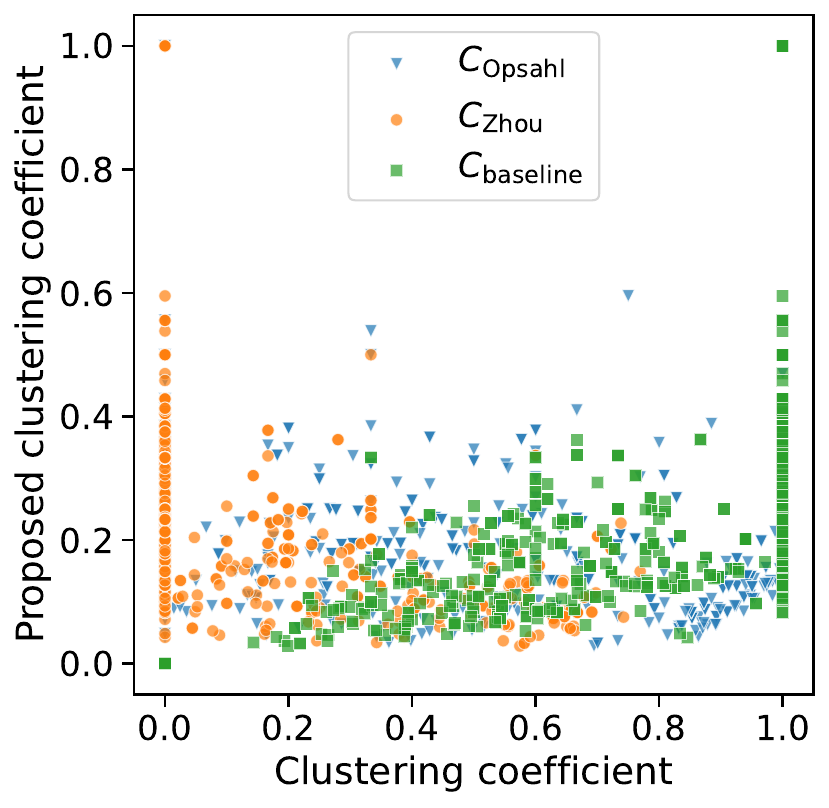}\\$\rho_{\mathrm{OP}}=-0.359$, $\rho_{\mathrm{ZP}}=-0.408$, \\$\rho_\mathrm{SP}=0.599$.\end{minipage}\label{fig:2c}}
    \caption{The histogram and scatter plot of the clustering coefficients.
    $\rho_{\mathrm{OP}}$, $\rho_{\mathrm{ZP}}$, and $\rho_\mathrm{SP}$ are Pearson's correlation coefficients of $C_{\mathrm{Opsahl}}$ and $C_{\mathrm{proposed}}$, $C_{\mathrm{Zhou}}$ and $C_{\mathrm{proposed}}$, $C_\mathrm{baseline}$ and $C_\mathrm{proposed}$, respectively.
    The proposed clustering coefficients $C_\mathrm{proposed}$ show most similar distribution to $C_\mathrm{Opsahl}$, $C_\mathrm{Zhou}$, and $C_\mathrm{baseline}$ on the primary-school dataset and different distributions on the other datasets.}
    \label{figure:dist}
\end{figure*}

\begin{table}[tp]
    \centering
    \caption{Number of nodes with clustering coefficient of 0 or 1 across different datasets.}
    \begin{tabular}{lrrrrrrrr}
    \toprule
    & \multicolumn{4}{c}{$\text{Clustering Coefficient} = 0$} & \multicolumn{4}{c}{$\text{Clustering Coefficient} = 1$} \\
    \cmidrule(r){2-5} \cmidrule(r){6-9}
    Dataset & $C_\mathrm{Opsahl}$ & $C_\mathrm{Zhou}$ & $C_\mathrm{proposed}$ & $C_\mathrm{baseline}$ & $C_\mathrm{Opsahl}$ & $C_\mathrm{Zhou}$ & $C_\mathrm{proposed}$ & $C_\mathrm{baseline}$ \\
    \midrule
    primary-school & 0 & 0 & 0 & 0 & 1 & 0 & 1 & 1 \\
    email-Enron & 2 & 2 & 1 & 1 & 4 & 1 & 1 & 4 \\
    NDC-classes & 283 & 385 & 17 & 17 & 9 & 1 & 5 & 303 \\
    \bottomrule
    \end{tabular}
    \label{tb:cc}
\end{table}
\end{DIFnomarkup}

The clustering coefficient is a metric that measures inherent clustering tendencies in hypergraphs.
Assuming that differences in clustering tendencies between datasets are greater than differences between definitions, our proposed definition should measure clustering tendencies in the same way as existing definitions.
To verify this, we calculated the clustering coefficients for each dataset.
The results are shown in \Cref{table:Dataset}.
All average clustering coefficients decrease in the same order: primary-school, email-Enron, and NDC-classes, indicating that the proposed definition effectively functions as an indicator of inherent clustering tendencies in hypergraphs.
The relative ordering of coefficients across the three datasets is consistent, suggesting that our proposed definition measures clustering tendencies similarly to existing definitions.
The baseline tends to yield larger values as hyperedge size increases because it assumes strong relationships between all nodes within a hyperedge, regardless of size.

\Cref{figure:dist} presents the distribution of clustering coefficients across all datasets. In the primary-school dataset, our proposed clustering coefficient exhibits a distribution pattern similar to other metrics, particularly aligning closely with the baseline ($\rho_{SP} = 0.998$). This similarity can be attributed to the prevalence of size-2 hyperedges in this dataset.
The email-Enron dataset shows more pronounced differences between existing and proposed definitions compared to the primary-school dataset. Notably, the correlation coefficient between Zhou's definition and our proposed definition is quite low at 0.346.
The most significant divergence between our proposed definition and existing approaches appears in the NDC-classes dataset. \Cref{fig:2c} reveals that existing definitions and the baseline frequently assign extreme values (0 or 1) to many nodes.
\Cref{tb:cc} quantifies this observation by showing the number of nodes assigned clustering coefficients of either 0 or 1 under each definition. These results demonstrate that our proposed definition assigns fewer extreme values. This is particularly evident in the NDC-classes dataset, where existing definitions frequently calculate clustering coefficients of 0 for many nodes, while our proposed definition successfully computes non-zero values.
Conversely, $C_\mathrm{baseline}$ tends to assign the extreme value of 1 to numerous nodes in the NDC-classes dataset.

The differences in values between our proposed definition and existing definitions can be attributed to two main factors.
First, the datasets contain different distributions of motifs. The NDC-classes dataset contains a higher proportion of order-3 motifs III and IV (see \Cref{figure:Motifs}), which our proposed definition can properly evaluate. This results in more appropriate values for datasets with abundant occurrences of these particular motifs.
However, since hypergraphs also contain motifs of order 4 or higher, the differences between our approach and existing definitions cannot be fully explained by the distribution of order-3 motifs alone.
Second, the presence of large hyperedges impacts the calculations.
Existing definitions focus exclusively on relationships formed between three distinct hyperedges while ignoring the internal structure within each hyperedge, particularly the relationships between pairs of nodes within a single large hyperedge.
Our proposed definition, however, considers these intra-hyperedge relationships.
Consequently, when large hyperedges are present, the difference in calculated values between our definition and existing approaches becomes more pronounced.
For a detailed investigation of the relationship between hyperedge size and clustering coefficients, please refer to the Supplementary Information.
While traditional approaches can only calculate non-zero clustering coefficients when three hyperedges form a triangle, our definition can compute meaningful non-zero values whenever hyperedges representing relationships between three or more nodes exist.
Therefore, our proposed clustering coefficient is particularly well-suited for datasets with numerous large hyperedges but lack small hyperedges, such as those found in social community networks and collaboration networks, where nodes simultaneously contained within a single hyperedge are assumed to have certain relationships.

\section*{Conclusion}

We proposed a novel clustering coefficient definition for hypergraphs that captures local link density by utilizing pairwise relationships within hyperedges.
Our approach transforms hypergraphs into weighted undirected graphs, where edge weights reflect connection strength based on hyperedge sizes, then calculates the local clustering coefficient on the resulting graph.
This approach enables more detailed reflection of pairwise node relationships compared to existing hypergraph clustering coefficient definitions.
Our theoretical evaluation on higher-order motifs of order 3 demonstrated that the proposed clustering coefficient satisfies three key conditions: (1) values fall within the range [0, 1], (2) consistency with clustering coefficients for undirected simple graphs, and (3) effective capture of pairwise relationships within hyperedges. Notably, our definition assigns meaningful non-zero values to motifs III, IV-a, and IV-b of order 3, where existing definitions fail.
The empirical evaluation on three real-world hypergraph datasets---primary-school, email-Enron, and NDC-classes---further validated our approach. The proposed clustering coefficient successfully measured inherent clustering tendencies similar to existing definitions while avoiding extreme values (0 or 1) in cases where they would be inappropriate. This was particularly evident in the NDC-classes dataset, where our definition calculated non-zero values for many nodes that would receive a clustering coefficient of 0 under existing definitions.

Several promising avenues for future research emerge from this work:
\begin{itemize}
\item Extension to weighted and directed hypergraphs: Our current definition is specifically designed for unweighted and undirected hypergraphs. Extending this concept to weighted or directed hypergraphs would broaden its applicability to a wider range of complex network representations, particularly in domains where edge direction and weight are critical factors.
\item Dynamic hypergraph analysis: Investigating how the proposed clustering coefficient changes over time in evolving hypergraphs could provide insights into the temporal dynamics of complex systems and their higher-order interactions.
\item Theoretical connections to other hypergraph metrics: Exploring the relationships between our proposed clustering coefficient and other hypergraph centrality measures could lead to a more comprehensive framework for hypergraph analysis.
\item Investigation of alternative weight functions: While our current weight function effectively captures pairwise relationships, exploring other theoretically motivated weight functions that satisfy our three key requirements could provide additional insights into the structure of hypergraphs and potentially reveal different aspects of local clustering patterns.
\end{itemize}
Future work could explore the application of the proposed clustering coefficient in various domains where complex hypergraphs naturally arise, such as collaboration networks, cellular networks, and social networks, potentially revealing new structural insights unique to each application domain.

\section*{Data availability}

The datasets used in this study are publicly available at \url{https://github.com/arbenson/ScHoLP-Data} by the original authors\cite{Benson2018}.
Codes to generate the results of the paper are available at \url{https://github.com/shudolab/hypergcc}.

\bibliography{hypercc}

\begin{thebibliography}{10}
\urlstyle{rm}
\expandafter\ifx\csname url\endcsname\relax
  \def\url#1{\texttt{#1}}\fi
\expandafter\ifx\csname urlprefix\endcsname\relax\def\urlprefix{URL }\fi
\expandafter\ifx\csname doiprefix\endcsname\relax\def\doiprefix{DOI: }\fi
\providecommand{\bibinfo}[2]{#2}
\providecommand{\eprint}[2][]{\url{#2}}

\bibitem{Klimt2004}
\bibinfo{author}{Klimt, B.} \& \bibinfo{author}{Yang, Y.}
\newblock \bibinfo{title}{The enron corpus: A new dataset for email
  classification research}.
\newblock In \emph{\bibinfo{booktitle}{European conference on machine
  learning}}, \bibinfo{pages}{217--226} (\bibinfo{year}{2004}).

\bibitem{Aksoy2020}
\bibinfo{author}{Aksoy, S.~G.}, \bibinfo{author}{Joslyn, C.},
  \bibinfo{author}{Marrero, C.~O.}, \bibinfo{author}{Praggastis, B.} \&
  \bibinfo{author}{Purvine, E.}
\newblock \bibinfo{journal}{\bibinfo{title}{Hypernetwork science via high-order
  hypergraph walks}}.
\newblock {\emph{\JournalTitle{EPJ Data Science}}}
  \textbf{\bibinfo{volume}{9}}, \bibinfo{pages}{16} (\bibinfo{year}{2020}).

\bibitem{Zhou2011}
\bibinfo{author}{Zhou, W.} \& \bibinfo{author}{Nakhleh, L.}
\newblock \bibinfo{journal}{\bibinfo{title}{Properties of metabolic graphs:
  biological organization or representation artifacts?}}
\newblock {\emph{\JournalTitle{BMC Bioinformatics}}}
  \textbf{\bibinfo{volume}{12}}, \bibinfo{pages}{132} (\bibinfo{year}{2011}).

\bibitem{Gallagher2013}
\bibinfo{author}{Gallagher, S.~R.} \& \bibinfo{author}{Goldberg, D.~S.}
\newblock \bibinfo{title}{Clustering coefficients in protein interaction
  hypernetworks}.
\newblock In \emph{\bibinfo{booktitle}{Proceedings of the International
  Conference on Bioinformatics, Computational Biology and Biomedical
  Informatics}}, \bibinfo{pages}{552--560} (\bibinfo{year}{2013}).

\bibitem{Yang2019}
\bibinfo{author}{Yang, D.}, \bibinfo{author}{Qu, B.}, \bibinfo{author}{Yang,
  J.} \& \bibinfo{author}{Cudre-Mauroux, P.}
\newblock \bibinfo{title}{Revisiting user mobility and social relationships in
  {LBSNs}: {A} hypergraph embedding approach}.
\newblock In \emph{\bibinfo{booktitle}{The {World} {Wide} {Web} {Conference}}},
  \bibinfo{pages}{2147--2157} (\bibinfo{year}{2019}).

\bibitem{Cattuto2007}
\bibinfo{author}{Cattuto, C.} \emph{et~al.}
\newblock \bibinfo{journal}{\bibinfo{title}{Network {Properties} of
  {Folksonomies}}}.
\newblock {\emph{\JournalTitle{AI Communications}}}
  \textbf{\bibinfo{volume}{20}}, \bibinfo{pages}{245--262}
  (\bibinfo{year}{2007}).

\bibitem{Zhang2010}
\bibinfo{author}{Zhang, Z.-K.} \& \bibinfo{author}{Liu, C.}
\newblock \bibinfo{journal}{\bibinfo{title}{A hypergraph model of social
  tagging networks}}.
\newblock {\emph{\JournalTitle{Journal of Statistical Mechanics: Theory and
  Experiment}}} \textbf{\bibinfo{volume}{2010}}, \bibinfo{pages}{P10005}
  (\bibinfo{year}{2010}).

\bibitem{Watts1998}
\bibinfo{author}{Watts, D.~J.} \& \bibinfo{author}{Strogatz, S.~H.}
\newblock \bibinfo{journal}{\bibinfo{title}{Collective dynamics of
  `small-world' networks}}.
\newblock {\emph{\JournalTitle{Nature}}} \textbf{\bibinfo{volume}{393}},
  \bibinfo{pages}{440--442} (\bibinfo{year}{1998}).

\bibitem{Masuda2018}
\bibinfo{author}{Masuda, N.}, \bibinfo{author}{Sakaki, M.},
  \bibinfo{author}{Ezaki, T.} \& \bibinfo{author}{Watanabe, T.}
\newblock \bibinfo{journal}{\bibinfo{title}{Clustering coefficients for
  correlation networks}}.
\newblock {\emph{\JournalTitle{Frontiers in Neuroinformatics}}}
  \textbf{\bibinfo{volume}{12}} (\bibinfo{year}{2018}).

\bibitem{Inoue2022}
\bibinfo{author}{Inoue, M.}, \bibinfo{author}{Pham, T.} \&
  \bibinfo{author}{Shimodaira, H.}
\newblock \bibinfo{journal}{\bibinfo{title}{A hypergraph approach for
  estimating growth mechanisms of complex networks}}.
\newblock {\emph{\JournalTitle{IEEE Access}}} \textbf{\bibinfo{volume}{10}},
  \bibinfo{pages}{35012--35025} (\bibinfo{year}{2022}).

\bibitem{Behague2023}
\bibinfo{author}{Behague, N.~C.}, \bibinfo{author}{Bonato, A.},
  \bibinfo{author}{Huggan, M.~A.}, \bibinfo{author}{Malik, R.} \&
  \bibinfo{author}{Marbach, T.~G.}
\newblock \bibinfo{journal}{\bibinfo{title}{The iterated local transitivity
  model for hypergraphs}}.
\newblock {\emph{\JournalTitle{Discrete Applied Mathematics}}}
  \textbf{\bibinfo{volume}{337}}, \bibinfo{pages}{106--119}
  (\bibinfo{year}{2023}).

\bibitem{Wu2016}
\bibinfo{author}{Wu, Z.}, \bibinfo{author}{Lin, Y.}, \bibinfo{author}{Wang, J.}
  \& \bibinfo{author}{Gregory, S.}
\newblock \bibinfo{journal}{\bibinfo{title}{Link prediction with node
  clustering coefficient}}.
\newblock {\emph{\JournalTitle{Physica A: Statistical Mechanics and its
  Applications}}} \textbf{\bibinfo{volume}{452}}, \bibinfo{pages}{1--8}
  (\bibinfo{year}{2016}).

\bibitem{Chen2019}
\bibinfo{author}{Chen, X.} \emph{et~al.}
\newblock \bibinfo{journal}{\bibinfo{title}{The application of degree related
  clustering coefficient in estimating the link predictability and predicting
  missing links of networks}}.
\newblock {\emph{\JournalTitle{Chaos: An Interdisciplinary Journal of Nonlinear
  Science}}} \textbf{\bibinfo{volume}{29}}, \bibinfo{pages}{053135}
  (\bibinfo{year}{2019}).

\bibitem{Estrada2006}
\bibinfo{author}{Estrada, E.} \&
  \bibinfo{author}{Rodr^^c3^^adguez-Vel^^c3^^a1zquez, J.~A.}
\newblock \bibinfo{journal}{\bibinfo{title}{Subgraph centrality and clustering
  in complex hyper-networks}}.
\newblock {\emph{\JournalTitle{Physica A: Statistical Mechanics and its
  Applications}}} \textbf{\bibinfo{volume}{364}}, \bibinfo{pages}{581--594}
  (\bibinfo{year}{2006}).

\bibitem{Kim2023}
\bibinfo{author}{Kim, S.}, \bibinfo{author}{Bu, F.}, \bibinfo{author}{Choe,
  M.}, \bibinfo{author}{Yoo, J.} \& \bibinfo{author}{Shin, K.}
\newblock \bibinfo{title}{How transitive are real-world group interactions? -
  {Measurement} and reproduction}.
\newblock In \emph{\bibinfo{booktitle}{Proceedings of the 29th ACM SIGKDD
  Conference on Knowledge Discovery and Data Mining}},
  \bibinfo{pages}{1132--1143} (\bibinfo{year}{2023}).

\bibitem{Ha2024}
\bibinfo{author}{Ha, G.-G.}, \bibinfo{author}{Neri, I.} \&
  \bibinfo{author}{Annibale, A.}
\newblock \bibinfo{journal}{\bibinfo{title}{Clustering coefficients for
  networks with higher order interactions}}.
\newblock {\emph{\JournalTitle{Chaos: An Interdisciplinary Journal of Nonlinear
  Science}}} \textbf{\bibinfo{volume}{34}}, \bibinfo{pages}{043102}
  (\bibinfo{year}{2024}).

\bibitem{Opsahl2013}
\bibinfo{author}{Opsahl, T.}
\newblock \bibinfo{journal}{\bibinfo{title}{Triadic closure in two-mode
  networks: {Redefining} the global and local clustering coefficients}}.
\newblock {\emph{\JournalTitle{Social networks}}}
  \textbf{\bibinfo{volume}{35}}, \bibinfo{pages}{159--167}
  (\bibinfo{year}{2013}).

\bibitem{Newman2018}
\bibinfo{author}{Newman, M.}
\newblock \emph{\bibinfo{title}{Networks}} (\bibinfo{publisher}{Oxford
  University Press}, \bibinfo{year}{2018}), \bibinfo{edition}{2} edn.

\bibitem{Robins2004}
\bibinfo{author}{Robins, G.} \& \bibinfo{author}{Alexander, M.}
\newblock \bibinfo{journal}{\bibinfo{title}{Small worlds among interlocking
  directors: Network structure and distance in bipartite graphs}}.
\newblock {\emph{\JournalTitle{Computational \& Mathematical Organization
  Theory}}} \textbf{\bibinfo{volume}{10}}, \bibinfo{pages}{69--94}
  (\bibinfo{year}{2004}).

\bibitem{Lind2005}
\bibinfo{author}{Lind, P.~G.}, \bibinfo{author}{Gonz^^c3^^a1lez, M.~C.} \&
  \bibinfo{author}{Herrmann, H.~J.}
\newblock \bibinfo{journal}{\bibinfo{title}{Cycles and clustering in bipartite
  networks}}.
\newblock {\emph{\JournalTitle{Physical Review E}}}
  \textbf{\bibinfo{volume}{72}}, \bibinfo{pages}{056127}
  (\bibinfo{year}{2005}).

\bibitem{Zhang2008}
\bibinfo{author}{Zhang, P.} \emph{et~al.}
\newblock \bibinfo{journal}{\bibinfo{title}{Clustering coefficient and
  community structure of bipartite networks}}.
\newblock {\emph{\JournalTitle{Physica A: Statistical Mechanics and its
  Applications}}} \textbf{\bibinfo{volume}{387}}, \bibinfo{pages}{6869--6875}
  (\bibinfo{year}{2008}).

\bibitem{Aksoy2017}
\bibinfo{author}{Aksoy, S.~G.}, \bibinfo{author}{Kolda, T.~G.} \&
  \bibinfo{author}{Pinar, A.}
\newblock \bibinfo{journal}{\bibinfo{title}{Measuring and modeling bipartite
  graphs with community structure}}.
\newblock {\emph{\JournalTitle{Journal of Complex Networks}}}
  \textbf{\bibinfo{volume}{5}}, \bibinfo{pages}{581--603}
  (\bibinfo{year}{2017}).

\bibitem{Latapy2008}
\bibinfo{author}{Latapy, M.}, \bibinfo{author}{Magnien, C.} \&
  \bibinfo{author}{Del~Vecchio, N.}
\newblock \bibinfo{journal}{\bibinfo{title}{Basic notions for the analysis of
  large two-mode networks}}.
\newblock {\emph{\JournalTitle{Social networks}}}
  \textbf{\bibinfo{volume}{30}}, \bibinfo{pages}{31--48}
  (\bibinfo{year}{2008}).

\bibitem{Zhou2006}
\bibinfo{author}{Zhou, D.}, \bibinfo{author}{Huang, J.} \&
  \bibinfo{author}{Sch^^c3^^b6lkopf, B.}
\newblock \bibinfo{title}{Learning with {Hypergraphs}: {Clustering},
  {Classification}, and {Embedding}}.
\newblock In \emph{\bibinfo{booktitle}{Advances in {Neural} {Information}
  {Processing} {Systems}}} (\bibinfo{year}{2006}).

\bibitem{Agarwal2006}
\bibinfo{author}{Agarwal, S.}, \bibinfo{author}{Branson, K.} \&
  \bibinfo{author}{Belongie, S.}
\newblock \bibinfo{title}{Higher order learning with graphs}.
\newblock In \emph{\bibinfo{booktitle}{Proceedings of the 23rd international
  conference on {Machine} learning}}, \bibinfo{pages}{17--24}
  (\bibinfo{year}{2006}).

\bibitem{Sheikhpour2025}
\bibinfo{author}{Sheikhpour, R.}, \bibinfo{author}{Berahmand, K.},
  \bibinfo{author}{Mohammadi, M.} \& \bibinfo{author}{Khosravi, H.}
\newblock \bibinfo{journal}{\bibinfo{title}{Sparse feature selection using
  hypergraph {Laplacian}-based semi-supervised discriminant analysis}}.
\newblock {\emph{\JournalTitle{Pattern Recognition}}}
  \textbf{\bibinfo{volume}{157}}, \bibinfo{pages}{110882}
  (\bibinfo{year}{2025}).

\bibitem{Liu2021}
\bibinfo{author}{Liu, M.}, \bibinfo{author}{Veldt, N.}, \bibinfo{author}{Song,
  H.}, \bibinfo{author}{Li, P.} \& \bibinfo{author}{Gleich, D.~F.}
\newblock \bibinfo{title}{Strongly {Local} {Hypergraph} {Diffusions} for
  {Clustering} and {Semi}-supervised {Learning}}.
\newblock In \emph{\bibinfo{booktitle}{Proceedings of the {Web} {Conference}
  2021}}, \bibinfo{pages}{2092--2103} (\bibinfo{year}{2021}).

\bibitem{Zhang2005}
\bibinfo{author}{Zhang, B.} \& \bibinfo{author}{Horvath, S.}
\newblock \bibinfo{journal}{\bibinfo{title}{A general framework for weighted
  gene co-expression network analysis}}.
\newblock {\emph{\JournalTitle{Statistical applications in genetics and
  molecular biology}}} \textbf{\bibinfo{volume}{4}}, \bibinfo{pages}{17}
  (\bibinfo{year}{2005}).

\bibitem{Lotito2022}
\bibinfo{author}{Lotito, Q.~F.}, \bibinfo{author}{Musciotto, F.},
  \bibinfo{author}{Montresor, A.} \& \bibinfo{author}{Battiston, F.}
\newblock \bibinfo{journal}{\bibinfo{title}{Higher-order motif analysis in
  hypergraphs}}.
\newblock {\emph{\JournalTitle{Communications Physics}}}
  \textbf{\bibinfo{volume}{5}}, \bibinfo{pages}{79} (\bibinfo{year}{2022}).

\bibitem{Benson2018}
\bibinfo{author}{Benson, A.~R.}, \bibinfo{author}{Abebe, R.},
  \bibinfo{author}{Schaub, M.~T.}, \bibinfo{author}{Jadbabaie, A.} \&
  \bibinfo{author}{Kleinberg, J.}
\newblock \bibinfo{journal}{\bibinfo{title}{Simplicial closure and higher-order
  link prediction}}.
\newblock {\emph{\JournalTitle{Proceedings of the National Academy of
  Sciences}}} \textbf{\bibinfo{volume}{115}}, \bibinfo{pages}{E11221--E11230}
  (\bibinfo{year}{2018}).

\bibitem{Stehle2011}
\bibinfo{author}{Stehl{\'e}, J.} \emph{et~al.}
\newblock \bibinfo{journal}{\bibinfo{title}{High-resolution measurements of
  face-to-face contact patterns in a primary school}}.
\newblock {\emph{\JournalTitle{PloS one}}} \textbf{\bibinfo{volume}{6}},
  \bibinfo{pages}{e23176} (\bibinfo{year}{2011}).

\end{thebibliography}

\section*{Acknowledgements}

This work was supported by JSPS KAKENHI Grant Number JP21H04872 and JP24H00691.
We would like to thank Dr. Kazuki Nakajima for useful discussions.

\section*{Author contributions statement}

R.M. and K.S. designed the work.
R.M. and S.H. conducted the experiments, and all authors analyzed the results.
R.M. and S.H. wrote the manuscript.
All authors reviewed the manuscript.

\section*{Additional information}

\subsection*{Competing interests}
The authors declare no competing interests.

\end{document}